\documentclass[aps,reprint,superscriptaddress,nofootinbib,twocolumn]{revtex4-2}
\usepackage[utf8]{inputenc}\usepackage[T1]{fontenc}
\usepackage{amsmath}
\usepackage{amssymb}
\usepackage{graphicx}
\usepackage{slashed}
\usepackage{cases}
\usepackage[toc,page]{appendix}
\usepackage{empheq}
\usepackage[normalem]{ulem} 
\usepackage{xcolor} 
\usepackage{afterpage}
\usepackage{float}
\usepackage{bbold}
\usepackage{hypbmsec}
\usepackage[colorlinks = true,
            linkcolor = blue,
            urlcolor  = blue,
            citecolor = blue,
            anchorcolor = blue]{hyperref}
\allowdisplaybreaks
\usepackage[export]{adjustbox}

\makeatletter

\newcommand{\be}{\begin{equation}}
\newcommand{\ee}{\end{equation}}
\newcommand{\bea}{\begin{eqnarray}}
\newcommand{\eea}{\end{eqnarray}}

\begin{document}

\title{Gravitational Wave Pathway to Testable Leptogenesis}

\author{Arnab Dasgupta}
\email{arnabdasgupta@pitt.edu}
\affiliation{Pittsburgh Particle Physics, Astrophysics, and Cosmology Center, Department of Physics and Astronomy, University of Pittsburgh, Pittsburgh, PA 15206, USA}

\author{P. S. Bhupal Dev}
\email{bdev@wustl.edu}
\affiliation{Department of Physics and McDonnell Center for the Space Sciences, Washington University, St. Louis, MO 63130, USA}

\author{Anish Ghoshal}
\email{anish.ghoshal@fuw.edu.pl}
\affiliation{Institute of Theoretical Physics, Faculty of Physics, University of Warsaw, ul.  Pasteura 5, 02-093 Warsaw, Poland}

\author{Anupam Mazumdar}
\email{anupam.mazumdar@rug.nl}
\affiliation{Van Swinderen Institute, University of Groningen, 9747 AG Groningen, The Netherlands}

\begin{abstract}
We analyze the classically scale-invariant $B-L$ model in the context of resonant leptogenesis with the recently proposed mass-gain mechanism. The $B-L$ symmetry breaking in this scenario is associated with a strong first order phase transition that gives rise to detectable gravitational waves (GWs) via bubble collisions. The same $B-L$ symmetry breaking also gives Majorana mass to right-handed neutrinos inside the bubbles, and their  out of equilibrium decays can produce the observed baryon asymmetry of the Universe via leptogenesis. We show that the current LIGO-VIRGO limit  on stochastic GW background already excludes part of the $B-L$ parameter space, complementary to the collider searches for heavy $Z^{\prime}$ resonances. Moreover, future GW experiments like Einstein Telescope and Cosmic Explorer can effectively probe the parameter space of leptogenesis over a wide range of the $B-L$ symmetry-breaking scales and gauge coupling values.

\end{abstract}

\maketitle

\section{Introduction}

The advent of gravitational wave (GW) astronomy has opened up a new observational window into the early Universe. A particularly interesting example of early Universe phenomena that can be a stochastic source of GWs is cosmological phase transition~\cite{Witten:1984rs,Hogan:1986qda}. Its investigation may play a crucial role in understanding an array of puzzles spanning from the baryon asymmetry of the Universe to the quest for an ultraviolet completion of the Standard Model (SM).  Although the electroweak phase transition is not predicted to be of first order within the SM~\cite{Rummukainen:1998as}, there are many extensions of the SM that predict strong first order phase transitions (SFOPTs) with detectable GWs~\cite{Kamionkowski:1993fg,Apreda:2001tj,Apreda:2001us,Grojean:2006bp,Espinosa:2008kw,Ashoorioon:2009nf,Das:2009ue,Dorsch:2014qpa,Kakizaki:2015wua,Jinno:2015doa,Huber:2015znp,Leitao:2015fmj,Huang:2016odd,Jaeckel:2016jlh,Dev:2016feu,Jinno:2016knw,Chala:2016ykx,Hashino:2016xoj,Artymowski:2016tme,Vaskonen:2016yiu,Dorsch:2016nrg,Baldes:2017rcu,Beniwal:2017eik,Marzola:2017jzl,Iso:2017uuu,Kang:2017mkl,Chala:2018ari,Bruggisser:2018mrt,Croon:2018erz,Megias:2018sxv,Okada:2018xdh,Baldes:2018emh,Prokopec:2018tnq,Beniwal:2018hyi,Brdar:2018num,Marzo:2018nov,Breitbach:2018ddu,Croon:2018kqn,Baratella:2018pxi,Angelescu:2018dkk,Alves:2018jsw,Fairbairn:2019xog,Hasegawa:2019amx,Dev:2019njv,Okada:2020vvb}. 


In this regard, classically conformal or scale-invariant models~\cite{Meissner:2006zh} provide good examples for generating sizable GW signals~\cite{Jinno:2016knw, Ellis:2020nnr}. This happens due to the fact that the tree-level potential is flat due to scale-invariance and thermal corrections easily dominate and makes the phase transition strongly first order~\cite{Ghoshal:2020vud,Hambye:2018qjv}.\footnote{Scale invariance makes potentials flat helping also in achieving successful inflation~\cite{Ghoshal:2022hyc,Ghoshal:2022qxk} and leads to robust predictions for dark matter due to constrained relation between couplings in the parameter space of the model~\cite{Hambye:2018qjv,Barman:2021lot,Barman:2022njh}.} According to  Bardeen's argument~\cite{Bardeen:1995kv}, once the classical conformal invariance and its minimal violation by the quantum anomalies are imposed on the SM, it can be free from the quadratic divergences, and hence, can cure the gauge hierarchy problem. In this case, all the mass scales must be generated by dimensional transmutation using the Coleman-Weinberg mechanism~\cite{Coleman:1973jx}. This mechanism cannot be applied directly to the SM Higgs sector to generate the electroweak scale since the predicted Higgs mass turns out to be always less than that of the $W$ boson mass, which is experimentally excluded. However, there are phenomenologically viable models with additional scalar(s) (and/or dark sectors) where the mass scale comes from the breaking of the conformal invariance involving those fields~\cite{Hempfling:1996ht,Espinosa:2007qk,Chang:2007ki,Foot:2007as,Foot:2007ay,Foot:2007iy,Meissner:2006zh,Meissner:2008gj,Iso:2009ss,Iso:2009nw,Barman:2021lot}. 

On the other hand, the observed matter-antimatter asymmetry in the Universe is one of the puzzles of modern cosmology that requires a dynamical explanation of how the Universe ended up having created more matter than antimatter, or more baryons than antibaryons, also known as baryogenesis. Among several proposed mechanisms for baryogensis (see Ref.~\cite{Bodeker:2020ghk} for a review), a particularly attractive variant is leptogenesis~\cite{Fukugita:1986hr} which involves lepton number violating (LNV) particles, such as the right-handed neutrinos (RHNs), to decay out of equilibrium and create a lepton asymmetry which later on gets converted to baryon asymmetry via the sphalerons~\cite{Kuzmin:1985mm}. The same RHNs participate in the seesaw mechanism~\cite{Minkowski:1977sc, Mohapatra:1979ia, Yanagida:1979as, GellMann:1980vs, Glashow:1979nm, Schechter:1980gr} for generating light neutrino masses. For a review on leptogenesis, see e.g.~Ref.~\cite{Davidson:2008bu}.

Recent work on baryogenesis via relativistic wall velocity has been proposed in Refs.~\cite{Azatov:2021irb,Baldes:2021vyz}, where it was shown that for classically scale-invariant models the particle gains mass instantaneously and hence becomes heavy and non-relativistic -- also known as the {\it mass-gain} mechanism. Now, if the particle has a baryon (lepton) number violating coupling then its out of equilibrium decay can produce a baryon (lepton) asymmetry as it becomes non-relativistic. 
In this paper we implement the mass-gain mechanism in a classically conformal\footnote{We have used the two terms ``scale" invariance and ``conformal" invariance interchangeably in this paper since they are known to be classically equivalent in any four-dimensional unitary and
renormalizable field theory~\cite{Gross:1970tb, Callan:1970ze, Coleman:1970je}.} $B-L$ model to achieve testable leptogenesis predictions at laboratory frontiers as well and show its correlation with observable GW signals in current and future detectors. 

The paper is structured as follows. In section~\ref{sec:Model} we have described the model under study and the effective potential with temperature correction. In section \ref{sec:Nucleation} we have described the nucleation temperature and the relevant constraints required for successful phase transition. In section \ref{sec:Lepto} we have presented our analysis for leptogenesis in this scenario. In section \ref{sec:GW} we explore the possibility of GWs in the leptogenesis parameter space. And finally in section \ref{sec:conclusion} we have concluded our study.

\section{Model and Effective Potential}
\label{sec:Model}

We consider the conformal $B-L$ extension of the SM~\cite{Iso:2009ss,Iso:2009nw} with the gauge group 
$SU(3)_c\times SU(2)_L\times U(1)_Y\times U(1)_{B-L}$. Three generations of RHNs  $\nu_R^i$ ($i = 1,2,3$) are introduced for anomaly cancellation.
An additional complex scalar field $\Phi$, charged under $U(1)_{B-L}$ is needed to spontaneously break the $U(1)_{B-L}$ gauge symmetry, which generates the masses of the RHNs. The particle content of the model is listed in Table~\ref{table_matcon}.

\begin{table}[t]
\begin{tabular}{|c||c|c|c|c|}
\hline
& $SU(3)_c$&$SU(2)_L$& $U(1)_Y$& $U(1)_{B-L}$ \\ \hline\hline
$q_L^i$& {\bf{3}}&{\bf 2}&+1/6&+1/3\\
$u_R^i$& {\bf{3}}&{\bf 1}&+2/3&+1/3\\
$d_R^i$& {\bf{3}}&{\bf 1}&$-1/3$&+1/3\\ \hline
$l_L^i$& {\bf{1}}&{\bf 2}&+1/6&$-1$\\
$e_R^i$& {\bf{1}}&{\bf 1}&$-1$&$-1$\\
$\nu_R^i$& {\bf{1}}&{\bf 1}&0&$-1$\\ \hline
$H$& {\bf{1}}&{\bf 2}&$-1/2$&0\\
$\Phi$& {\bf{1}}&{\bf 1}&0&+2\\ \hline
\end{tabular}
\caption{Particle content of the classically scale-invariant $B-L$ model. }
\label{table_matcon}
\end{table}

The additional Yukawa interactions involving the RHNs are given by  
\begin{align}
\mathcal{L}_Y
&\supset -Y_D^{ij}\bar{\nu}_R^i H^\dag l_L^j-\frac{1}{2}Y_M^{i}\Phi\bar{\nu}_R^{ic}\nu_R^i+\text{H.c.} ,
\end{align}
where the first term gives the Dirac neutrino mass after electroweak symmetry breaking, while the second term generates the RHN Majorana mass term.   
One may assume the Yukawa coupling $Y_M^i$ to have a diagonal form without loss of generality.
Neutrino masses are generated by the usual seesaw mechanism~\cite{Minkowski:1977sc, Mohapatra:1979ia, Yanagida:1979as, GellMann:1980vs, Glashow:1979nm, Schechter:1980gr} after the scalars $H$ and $\Phi$ acquire their vacuum expectation values.

The scale-invariant scalar potential looks like:
\begin{align}
V(H,\Phi)
&= \lambda_H(H^\dag H)^2+\lambda(\Phi^\dag \Phi)^2-\lambda'(\Phi^\dag \Phi)(H^\dag H),
\end{align}
where one may notice the absence of the quadratic mass terms. Thus the symmetry breaking must occur radiatively. When the Yukawa coupling $Y_M$ is negligible compared to the $U(1)_{B-L}$ gauge coupling, the $\Phi$ sector is the same as the original Coleman-Weinberg potential~\cite{Coleman:1973jx}. 
We consider simultaneous breaking of electroweak and $B-L$ symmetries due to radiative corrections via the $\lambda'$ term, and study the effective potential for the $\Phi$ field. 

\subsection{Zero-temperature effective potential}
Before we go on to the finite-temperature corrections, let us write down the one-loop corrected zero-temperature effective potential for $\phi \equiv \sqrt{2}\mathrm{Re}(\Phi)$~\cite{Meissner:2008uw,Iso:2009nw}: 
\begin{align}
V_0(\phi,t)
&= \frac{1}{4}\lambda(t)G(t)^4\phi^4,
\end{align}
where $t = \log(\phi/\mu)$, with $\mu$ being the renormalization scale and
\begin{align}
G(t)
&= \exp\left[-\int_0^tdt'\gamma(t')\right],
\;\;\;
\gamma(t)
= -\frac{a_2}{32\pi^2}g_{B-L}(t)^2,
\end{align}
with $a_2=24$.
The gauge and self coupling strengths $\alpha_{B-L}\equiv g_{B-L}^2/4\pi$ and 
$\alpha_\lambda \equiv \lambda/4\pi$ evolve according to the 
renormalization group equations (RGEs) as stated below:
\begin{align}
&2\pi \frac{d\alpha_{B-L}(t)}{dt}
= b\alpha_{B-L}(t)^2, \\
&2\pi \frac{d\alpha_\lambda(t)}{dt}
= a_1\alpha_\lambda(t)^2+8\pi\alpha_\lambda(t) \gamma(t)+a_3\alpha_{B-L}(t)^2,
\end{align}
with $b=12$, $a_1=10$, and $a_3=48$.
For the renormalization scale $\mu = M$, the stationary condition 
$\frac{dV}{d\phi}\big|_{\phi=M} = 0$ leads to a relation among the coupling constants as
\begin{align}
a_1\alpha_\lambda(0)^2 + a_3\alpha_{B-L}(0)^2 + 8\pi\alpha_\lambda(0)
&= 0,
\label{eq_cond_vev}
\end{align}
which means that $\alpha_{\lambda}(0)$ is determined by $\alpha_{B-L}(0)$; i.e. we have only two independent parameters, $M$ and $\alpha_{B-L}(0)$.
Analytical form of the scalar potential after including the RGE-improved form looks like:~\cite{Iso:2009ss}
\begin{align}
V_0(\phi,t)
&=\frac{\pi\alpha_\lambda(t)}{\left(1-\frac{b}{2\pi}\alpha_{B-L}(0)t\right)^{a_2/b}} \phi^4,\
\label{eq_V0}
\end{align}
where
\begin{align}
&\alpha_{B-L}(t)
= \frac{\alpha_{B-L}(0)}{1-\frac{b}{2\pi}\alpha_{B-L}(0)t},
\label{eq_aBLSolution}
\\
&\alpha_\lambda(t)
= \frac{a_2+b}{2a_1}\alpha_{B-L}(t) \nonumber 
\\
&\;\;\;\;\;\;\;\;\;\;\;\;\;
+\frac{A}{a_1}\alpha_{B-L}(t)\tan\left[
\frac{A}{b}\ln\left[\alpha_{B-L}(t)/\pi \right]+C
\right].
\end{align}
Here $A \equiv \sqrt{a_1a_3-(a_2+b)^2/4}$,
and the coefficient $C$ is determined such that Eq.~(\ref{eq_cond_vev}) is always true.

\subsection{Finite-temperature effective potential}

Let us define the renormalization scale parameter $u$ instead of $t$ as
\begin{align}
u
&\equiv \log(\Lambda/M), 
\label{eq_uDef}
\end{align}
where
\begin{align}	
\Lambda
&\equiv
\max (\phi,T)
\end{align}
represents the typical scale of the system considered, $T$ being the temperature. The one-loop level effective potential becomes:
\begin{align}
V_{\rm eff}(\phi,T)
&= V_0(\phi,u)+V_T(\phi,T).
\label{eq_effpot}
\end{align}
Here $V_0$ indicates the zero-temperature potential (\ref{eq_V0}), while $V_T$ denotes the thermal corrections:
\begin{align}
V_T(\phi,T)
&=\frac{3}{2}V_T^B(m_V(\phi)/T,T)+V_{\rm daisy}(\phi,T),
\end{align}
where
\begin{align}
V_T^B(x,T)
&\equiv \frac{T^4}{\pi^2}\int_0^\infty dz~z^2\ln \left[ 1-e^{-\sqrt{z^2+x^2}} \right],
\end{align}
is the bosonic one-loop contribution, and
\begin{align}
V_{\rm daisy}(\phi,T)
&= -\frac{T}{12\pi}\left[m_V^3(\phi,T) - m_V^3(\phi)\right]
\end{align}
is the so-called daisy subtraction~\cite{Arnold:1992rz}.
The thermal mass of the $B-L$ vector boson is given by
\begin{align}
	m_V^2(\phi,T)&=m_V^2(\phi)+c_t g_{B-L}^2(t)T^2,
\end{align}
where $m_V(\phi)=2g_{B-L}(t)\phi$, and $c_t=4$. Here the self-interaction contribution of $\phi$ 
to the thermal potential is not taken into account, since it is much smaller than the one from the gauge interaction.

\section{Nucleation Temperature and feasiblity of successful Phase transition}
\label{sec:Nucleation}



Let us approximate the effective potential with the dominant temperature contributions:
\begin{align}
V_{\rm eff}
&\simeq \frac{g_{B-L}^2(u)}{2}T^2\phi^2 + \frac{\lambda_{\rm eff}(u)}{4}\phi^4,
\label{eq_Veff_approx}
\end{align}
with
$\lambda_{\rm eff}(u) \equiv 4\pi\alpha_\lambda(u) / \left( 1-\frac{b}{2\pi}\alpha_{B-L}(0)u \right)^{a_2/b}$
(see Eq.~(\ref{eq_V0})).
For $T \gg M$, 
the effective potential has a unique minimum at $\phi = 0$, and for $T \ll M$, 
 $\phi=0$ becomes a false vacuum point.

\begin{figure}
    \centering
    \includegraphics[width=0.45\textwidth]{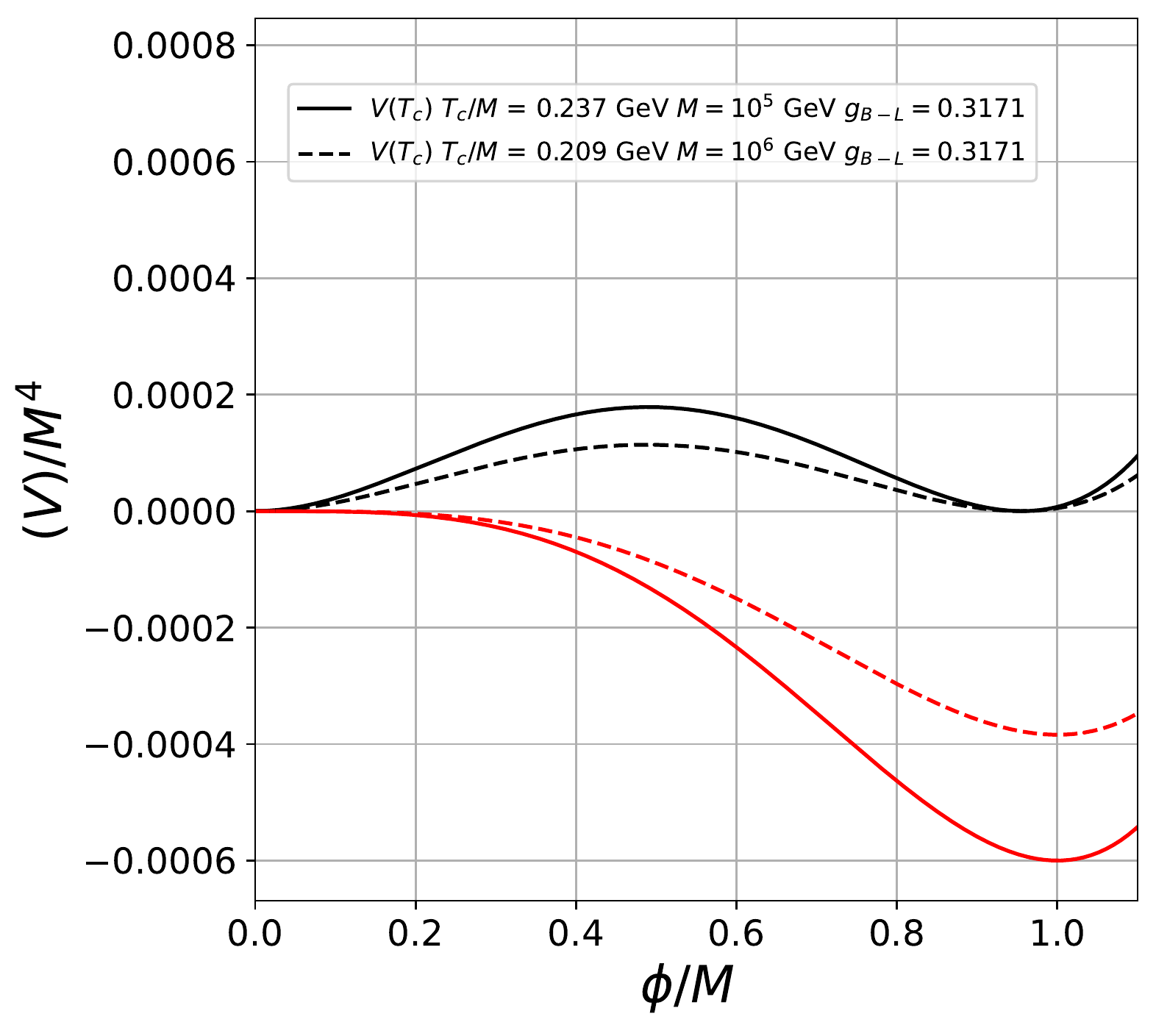}
    \caption{Evolution of the effective potential for two benchmark points (solid and dashed). The black (red) curve corresponds to temperature $T$ at (below) the critical temperature $T_c$ for $B-L$ breaking scale.}
    \label{fig:potential}
\end{figure}

In Fig.~\ref{fig:potential}, we show the evolution of the effective potential for two benchmark points (solid and dashed). The black (red) curve corresponds to temperature $T$ at (below) the critical temperature $T_c$ for $B-L$ breaking scale. The $\phi$ field is initially trapped at the origin of the effective potential and then as temperature drops below $T = T_c \sim M$, the Universe experiences a phase transition 
associated with the tunneling of the $\phi$ field from false to true vacuum triggering bubble nucleation and subsequent GW production. This phase transition is first order provided the transition rate exceeds the expansion rate of the Universe. %

\subsection{Nucleation rate}

The nucleation rate per unit volume $\Gamma$ is 
given by~\cite{Linde:1977mm,Linde:1981zj}
\begin{align}
&\Gamma (T)
= A(T) e^{-S_3(T)/T},
\label{eq_Gamma1}
\end{align}
with the three-dimensional action 
\begin{align}
&S_3(T)
= \int d^3x \left[ \frac{1}{2} (\nabla \phi)^2 + \left( V_{\rm eff}(\phi,T) - V_{\rm eff}(0,T) \right) \right] .
\label{eq_S3}
\end{align}
In Eq.~\eqref{eq_Gamma1}, $A$ denotes a prefactor which is typically of $\mathcal{O}(T^4)$.\footnote{
We consider $g_{B-L}\lesssim 0.3$ in the following. 
In such a case, the effects from the prefactor are negligible 
(see Eqs. (8) and (9) in Ref.~\cite{Strumia:1999fv}).
}
The configuration of $\phi$ in $S_3$ is 
estimated from
\begin{align}
\frac{d^2 \phi}{dr^2} + \frac{2}{r} \frac{d\phi}{dr} - \frac{\partial V_{\rm eff}}{\partial \phi}
&= 0 \, ,
\end{align}
with the following boundary conditions:
\begin{align}
\phi(r = \infty)
&= 
\phi_{\rm false}, 
\;\;\;\;\;\;
\frac{d\phi}{dr} (r=0)
= 0 \, .
\end{align}
Using Eq.~(\ref{eq_Gamma1}) we get the transition rate as
\begin{align}
\Gamma(T)
&=
Be^{-S(T)}, 
\label{eq_Gamma2} 
\end{align}
with
\begin{align}
B
&\equiv M^4 \frac{A(T)}{T^4} \, , \\
S(T)
&\equiv \frac{S_3(T)}{T} - 4\log(T/M) \, .
\label{eq_S}
\end{align}
Since $A$ is of ${\mathcal O}(T^4)$, the transition rate is dominantly determined by $S$\footnote{For comparison between O(3) vs O(4) bounce solutions, see Ref.~\cite{Ghoshal:2020vud}.}.

Now, considering the broken-phase regime $T\ll M$, the effective
potential around the origin $\phi \lesssim T$ is approximately given by Eq.~\eqref{eq_Veff_approx} with $u=t= \ln{(T/M)}$. In such a case, the action is given as~\cite{Linde:1981zj}
\begin{align}
    S &= \frac{S_3}{T} - 4\ln{(T/M)}, \nonumber \\
    \frac{S_3}{T} &\simeq - 19\times \frac{g_{B-L}(t)}{\lambda_{\rm eff}(t)} \, .
\end{align}

The nucleation temperature is defined as the inverse time of creation of one bubble per Hubble radius which is given as:
\begin{align*}
    \left.\Gamma/H^4\right|_{T=T_n} =1 \, .
\end{align*}

\subsection{Vacuum Transition Probability}
In this section we now go from one bubble treatment to statistical analysis of bubbles in the early Universe. The probability
for a given point to be in the unstable vacuum is given by $p(T) = e^{-I(T)}$, where~\cite{Ellis:2020nnr}
\begin{align}
    I(T) &= \frac{4\pi}{3}\int^{T_c}_{T}\frac{dT^\prime}{T^{\prime 4}}\frac{\Gamma(T^\prime)}{H(T^\prime)}\left(\int^{T^\prime}_{T}\frac{d\tilde{T}}{H(\tilde{T})}\right)^3.
\end{align}
Here $H(T)\simeq 17 \: T^2/M_{\rm Pl}$ is the Hubble expansion rate at temperature $T$ ($M_{\rm Pl}$ being the Planck mass). 
In order to calculate the percolation temperature, $T_p$, we solve the above integral with $I(T_p)=0.34$, which in other words implies that $34\%$ of the comoving volume has converted to the true minimum. In addition to this requirement, a stronger condition which needs to be satisfied is that the volume of the false vacuum should decrease i.e. 
\begin{align}
    \frac{1}{{V}_{\rm false}}\frac{d{V}_{\rm false}}{dt} &= H(T)\left(3+\frac{dI(T)}{dT}\right) < 0.
\end{align}

\section{Leptogenesis}
\label{sec:Lepto}
Our leptogenesis analysis closely follows the recent work of Ref.~\cite{Baldes:2021vyz}, i.e.~the {\it mass-gain} mechanism. We first need to ensure that the Lorentz boost of the bubble wall satisfies the following criterion: 
\begin{align}
\gamma_w  = \frac{1}{\sqrt{1-v^2_w}} > M_N(T_n)/T_n  \, ,
\label{eq:gammaw}
\end{align}
where $T_n$ is the nucleation temperature and $M_N(T)$ is the thermal mass of the RHN: 
\begin{align}
    M^2_N(T) &=  Y^2_M M^2 + \frac{1}{8}g^2_{B-L}T^2 \, .
    \label{eq:therm-RHN}
\end{align}
The condition~\eqref{eq:gammaw} basically pushes the RHN quanta into the bubble while maintaining the equilibrium comoving number density 
\begin{align}
    Y_N &= \frac{135}{8\pi^4}\xi(3)\frac{g_N}{g_*} \, ,
\end{align}
where $g_N$ and $g_*$ are the degrees of freedom of $N$ and the relativistic degrees of freedom, respectively. In order to calculate the bubble wall velocity we first calculate the {\it Jouguet} velocity~\cite{Kamionkowski:1993fg,Steinhardt:1981ct,Espinosa:2010hh}
\begin{align}
    v_J &= \frac{1}{\sqrt{3}}\frac{1+\sqrt{3\alpha^2 + 2\alpha}}{1+ \alpha} \, ,
\end{align}
where $\alpha$ is change in the trace of the energy-momentum tensor, $\Delta T^\mu_\mu$, across the phase transition~\cite{Caprini:2019egz},
\begin{align}
    \alpha &= \left.\frac{1}{\rho_r}\left[\Delta V - \frac{T}{4}\frac{d\Delta V}{dT} \right]\right|_{T=T_n} \, ,
    \label{eq:alpha}
\end{align}
and $\Delta V$ is the potential difference between the true and false vacuum and $\rho_r$ is the radiation energy density. 
The rough estimate of the bubble wall velocity is then given as~\cite{Lewicki:2021pgr}
\begin{align}
    v_w = \begin{cases} \sqrt{\frac{\Delta V}{\alpha \rho_r}} & \quad {\rm for} \quad \sqrt{\frac{\Delta V}{\alpha \rho_r}} < v_J \\ 
    1 & \quad {\rm for} \quad \sqrt{\frac{\Delta V}{\alpha \rho_r}} \geq v_J\end{cases} .
\end{align}
As we will see later (cf.~Table~\ref{tab:BP}), for the choice of our benchmark points, the second condition is always satisfied, i.e. $\sqrt{\frac{\Delta V}{\alpha \rho_r}} \geq v_J$, and the wall velocity is always equal to 1; therefore, $\gamma_w$ is infinity and the condition \eqref{eq:gammaw} is trivially satisfied.\footnote{This is a special feature of the mass-gain mechanism, in contrast with the conventional electroweak baryogenesis scenarios, where the baryon asymmetry goes to zero in the limit of $v_w\to 1$~\cite{Cline:2020jre}.}

Now, the RHNs are already out of equilibrium within the bubble and thus can decay via CP-violating processes $N\rightarrow L H, \bar{L}H^\dag$ to generate a leptonic asymmetry~\cite{Fukugita:1986hr} that gets transformed into a baryonic asymmetry via the electroweak sphalerons~\cite{Kuzmin:1985mm}. The final baryonic asymmetry can be written as follows:
\begin{align}
    Y_{B} = \epsilon_N \kappa_{\rm Sph}Y_N\left(\frac{T_n}{T_{\rm RH}}\right)^3.
    \label{eq:eps}
\end{align}
where $\epsilon_N$ is the CP-asymmetry in the decay of RHNs, $\kappa_{\rm Sph} = 28/79$ is the sphaleron conversion rate, and $T_{\rm RH}$ is the reheating temperature. The $Y_B$ obtained in Eq.~\eqref{eq:eps} should then be compared with the observed baryon asymmetry normalized over the entropy density: $Y_B^{\rm obs}=(8.61\pm 0.05)\times 10^{-11}$~\cite{Planck:2018vyg}. 

In order to check the feasibility for the decay of the RHNs to Higgs and leptons, we need to first consider the thermally corrected masses for the Higgs and lepton doublets at the reheating temperature~\cite{Giudice:2003jh}:
\begin{align}
    M_H^2(T) &= \left(\frac{3}{16}g^2_2 + \frac{1}{16}g^2_Y + \frac{1}{4}y^2_t\right)T^2 \, , \nonumber \\
    M^2_L(T) &= \left(\frac{3}{32}g^2_2 + \frac{1}{32}g^2_Y\right) T^2  \, , 
\end{align}
where $g_Y$ and $g_2$ are the $U(1)_Y$ and $SU(2)_L$ gauge couplings respectively, and $y_t$ is the top Yukawa coupling. At the reheating temperature, we get \begin{align}
    M_H(T_{\rm RH}) + M_L(T_{\rm RH})&\simeq 0.77 T_{\rm RH} , 
    \label{eq:therm1}
\end{align}
where we have set the coupling values at the electroweak scale.\footnote{The values do not change much between the electroweak scale and the reheating temperature for (multi) TeV-scale symmetry breaking considered here.} 
The sum of thermal masses of the Higgs and lepton in Eq.~\eqref{eq:therm1}  needs to be lower than the RHN mass given by Eq.~\eqref{eq:therm-RHN} at $T=T_{\rm RH}$. 

Now there are two possibilities depending on the size of the Yukawa coupling:
\begin{enumerate}
    \item $Y_M M \leq 0.35 g_{B-L} T_{\rm RH}$ 
    \item $Y_M M > 0.35 g_{B-L} T_{\rm RH}$
\end{enumerate}
For the first possibility the lower limit for the $g_{B-L}$ is needed to be $g_{B-L} \geq 2.2$ for the RHN decay to be possible. Such large gauge couplings will hit the Landau pole very quickly, making the theory invalid at  a relatively low scale. On the other hand, if we have second possibility then the condition for the RHN Yukawa coupling will be 
\begin{align}
    Y_M \equiv y_f g_{B-L} \geq 0.77 \frac{T_{\rm RH}}{M}, \quad {\rm or},~
    y_f \geq \frac{0.77}{g_{B-L}} \frac{T_{\rm RH}}{M} \, .
    \label{eq:yf}
\end{align}
For concreteness, we consider a $y_f$ value twice the lower limit in Eq.~\eqref{eq:yf} to compute the lepton asymmetry from $N\to LH$ decay.  

Furthermore, we have considered the input from the neutrino mass matrix to constrain the Dirac Yukawa coupling $Y_D$ which is primarily responsible for the washout of the generated asymmetry. We have calculated the Dirac Yukawa coupling by considering the Casas-Ibarra parametrization~\cite{Casas:2001sr}

\begin{align}
    Y_D &= \Lambda^{-1/2}\mathcal{O}\widehat{m}_{\nu}^{1/2}U^\dagger_{\rm PMNS} \, ,
\end{align}
    where $\Lambda=v^2/M_N$, ${\cal O}$ is an arbitrary complex orthogonal matrix, $\widehat{m}_\nu$ is the diagonal light neutrino mass matrix and $U_{\rm PMNS}$ is the light neutrino mixing matrix. Using the best-fit values of the light neutrino oscillation data~\cite{Gonzalez-Garcia:2021dve} for normal hierarchy and assuming ${\cal O}$ to be the identity matrix, we obtain  
    \begin{align}
    y \equiv \sum_\alpha Y_{D_{1\alpha}} \sim  2.3\times 10^{-8}\left(\frac{M_N}{1 ~{\rm TeV} }\right)^{1/2} \, .
\end{align}
Since we are dealing with leptogenesis at energy scales below the so-called Davidson-Ibarra bound~\cite{Davidson:2002qv}, we invoke the resonant leptogenesis mechanism~\cite{Pilaftsis:2003gt}, where the dominant contribution to the CP asymmetry comes from the wave function corrections, and is independent of the size of the Yukawa couplings. However, for maximal CP asymmetry, it requires a fine-tuning in the mass difference between the two RHNs which should be comparable to the RHN decay width: $\Delta M/M \sim \Gamma/(2M) = y^2/(32\pi)$.\footnote{For our benchmark points in Table~\ref{tab:BP}, this amounts to a fine-tuning of one part in $10^{12}$ or so. However, there exist various symmetry-motivated mechanisms to generate such small mass splittings from an exactly degenerate RHN mass spectrum; see e.g.~Refs.~\cite{GonzalezFelipe:2003fi, Dev:2014pfm, Chauhan:2021xus, Drewes:2022kap}.} In this case, the CP-asymmetry in Eq.~\eqref{eq:eps} can be written as $\epsilon_N \simeq \sin(2\phi)/16\pi$, where $\phi$ is the relative CP phase between the two RHNs. In our analysis, we have chosen the phase $\phi$ in such a way that  the correct baryon asymmetry can always be obtained from Eq.~\eqref{eq:eps}.

But at reheating we require the washout process coming from the inverse decay $LH\to N$ to be out of equilibrium 
which is possible if the following condition is satisfied~\cite{Baldes:2021vyz}:
\begin{align}
    \frac{M_N}{T_{\rm RH}} \gtrsim \ln\left[\frac{y^2M_{\rm Pl}}{24\pi T_{\rm RH}}\left(\frac{M_N}{T_{\rm RH}}\right)^{5/2}\right] .
\end{align}
Considering $M_N \simeq 1.54 T_{\rm RH}$ we satisfy the above relation.

\section{Gravitational Waves} \label{sec:GW}
Before showing the correlation of the scale of leptogenesis with the present and future GW experiments let us first take a slight detour to understand the GW contributions coming from the SFOPT.
There are three main contributions to the GW amplitude coming from bubble collision ($\Omega_b$), sound wave ($\Omega_s$) and magnetohydrodynamic turbulance ($\Omega_t$). The linear superposition of these contributions gives the total GW amplitude: 
\begin{align}
    \Omega h^2 = \Omega_bh^2 + \Omega_sh^2 + \Omega_th^2 \, ,
\end{align}
where $h$ is the dimensionless Hubble parameter.  
Now each of these contributions relies on some basic parameters coming from the SFOPT, namely $\alpha$ (defined above in Eq.~\eqref{eq:alpha})
, $\beta/H_*$ (the inverse of the duration of the
phase transition in units of the Hubble time $H^{-1}_*$ at the time of GW production), $T_*$ (the characteristic temperature at the time of GW production), $v_w$ (bubble wall velocity), and $\kappa_{s,b,t}$ (the efficiency factors that characterize the fractions of the released vacuum energy that are converted into the energy of scalar-field gradients, sound waves, and turbulence, respectively). In terms of the peak amplitude, each contribution is given as follows~\cite{Schmitz:2020syl}\footnote{For bubbles in a gauge theory recent studies have shown that the GW spectrum maybe slightly modified from this general case, but the effect on the spectrum is tiny~\cite{Lewicki:2020azd}; so we only consider the general case, as usually done in the literature.}:
\begin{align}
    h^2\Omega_b(f) &= h^2\Omega_{\rm peak}(\alpha, \beta/H_*, T_*, v_w,\kappa_b) \mathcal{S}_b (f, f_b) \, ,  \nonumber \\ 
    h^2\Omega_s(f) &= h^2\Omega_{\rm peak}(\alpha, \beta/H_*, T_*, v_w,\kappa_s) \mathcal{S}_s (f, f_s) \, , \nonumber \\ 
    h^2\Omega_t(f) &= h^2\Omega_{\rm peak}(\alpha, \beta/H_*, T_*, v_w,\kappa_t) \mathcal{S}_t (f, f_t) \, , 
\end{align}
The peak of the amplitudes are given as~\cite{Schmitz:2020syl}:
\begin{widetext}
\begin{align}
    \Omega_{\rm peak}(\alpha, \beta/H_*, T_*, v_w,\kappa_b) &\simeq 1.67\times 10^{-5}\left(\frac{v_w}{\beta/H_*}\right)^2\left(\frac{100}{g_*(T_*)}\right)^{1/3}\left(\frac{\kappa_b\alpha}{1+\alpha}\right)^2\left(\frac{0.11v_w}{0.42 + v^2_w}\right) \, , \label{eq:bubble} \\
    \Omega_{\rm peak}(\alpha, \beta/H_*, T_*, v_w,\kappa_s) &\simeq 2.65\times 10^{-6}\left(\frac{v_w}{\beta/H_*}\right)\left(\frac{100}{g_*(T_*)}\right)^{1/3}\left(\frac{\kappa_s\alpha}{1+\alpha}\right)^2 \, , \label{eq:sound} \\
    \Omega_{\rm peak}(\alpha, \beta/H_*, T_*, v_w,\kappa_t) &\simeq 3.35\times 10^{-4}\left(\frac{v_w}{\beta/H_*}\right)\left(\frac{100}{g_*(T_*)}\right)^{1/3}\left(\frac{\kappa_t\alpha}{1+\alpha}\right)^{3/2} \, , \label{eq:mhd} 
    \end{align}
\end{widetext}
The corresponding peak frequencies are given by 
\begin{align}
    f_b &= 1.62\times 10^{-2}{\rm mHz}\left(\frac{g_*(T_*)}{100}\right)^{1/6}\left(\frac{T_*}{100 {\rm GeV}}\right)\left(\frac{\beta/H_*}{v_w}\right)\nonumber \\
    & \qquad \times \left(\frac{0.62v_w}{1.8 -0.1v_w + v^2_w}\right) \, , \\
    f_s &= 1.9\times 10^{-2}{\rm mHz}\left(\frac{g_*(T_*)}{100}\right)^{1/6}\left(\frac{T_*}{100 {\rm GeV}}\right)\left(\frac{\beta/H_*}{v_w}\right) \, , \\
    f_t &= 2.7\times 10^{-2}{\rm mHz}\left(\frac{g_*(T_*)}{100}\right)^{1/6}\left(\frac{T_*}{100 {\rm GeV}}\right)\left(\frac{\beta/H_*}{v_w}\right) \, .
\end{align}
The spectral shapes $\mathcal{S}$ are given as:
\begin{align}
    \mathcal{S}_b &= \left(\frac{f}{f_b}\right)^{2.8}\left(\frac{3.8}{1 + 2.8(f/f_b)^{3.8}}\right) \nonumber \\
    \mathcal{S}_s &= \left(\frac{f}{f_s}\right)^{3}\left(\frac{7}{4 + 3(f/f_s)^{2}}\right)^{7/2} \nonumber \\
    \mathcal{S}_t &= \left(\frac{f}{f_t}\right)^{3}\left(\frac{1}{1 + (f/f_t)}\right)^{11/3}\frac{1}{1+8\pi f/h_*}
\end{align}
where the Hubble frequency $h_*$ corresponds to the Hubble rate $H_*$ at the time of GW production. The redshifted value depends on $T_*$ as 
\begin{align}
    h_* &= 1.6\times 10^{-2} {\rm mHz} \left(\frac{g_*(T_*)}{100}\right)^{1/6}\left(\frac{T_*}{100 {\rm GeV}}\right).
\end{align}
Since the scenario we are particularly studying is the supercooled regime, the characteristic temperature is the reheating temperature $T_* = T_{\rm RH}$ as $T_n \ll T_{\rm RH}$. And the parameter $\beta/H_*$ in Eqs.~\eqref{eq:bubble}-\eqref{eq:mhd} is defined as:
\begin{align}
    \frac{\beta}{H_*} &= \frac{H_n}{H_*} T_n \left(\frac{dS}{dT}\right)_{T=T_n} \, .
\end{align}

\begin{figure}[!t]
    \centering
    \includegraphics[width=0.45\textwidth]{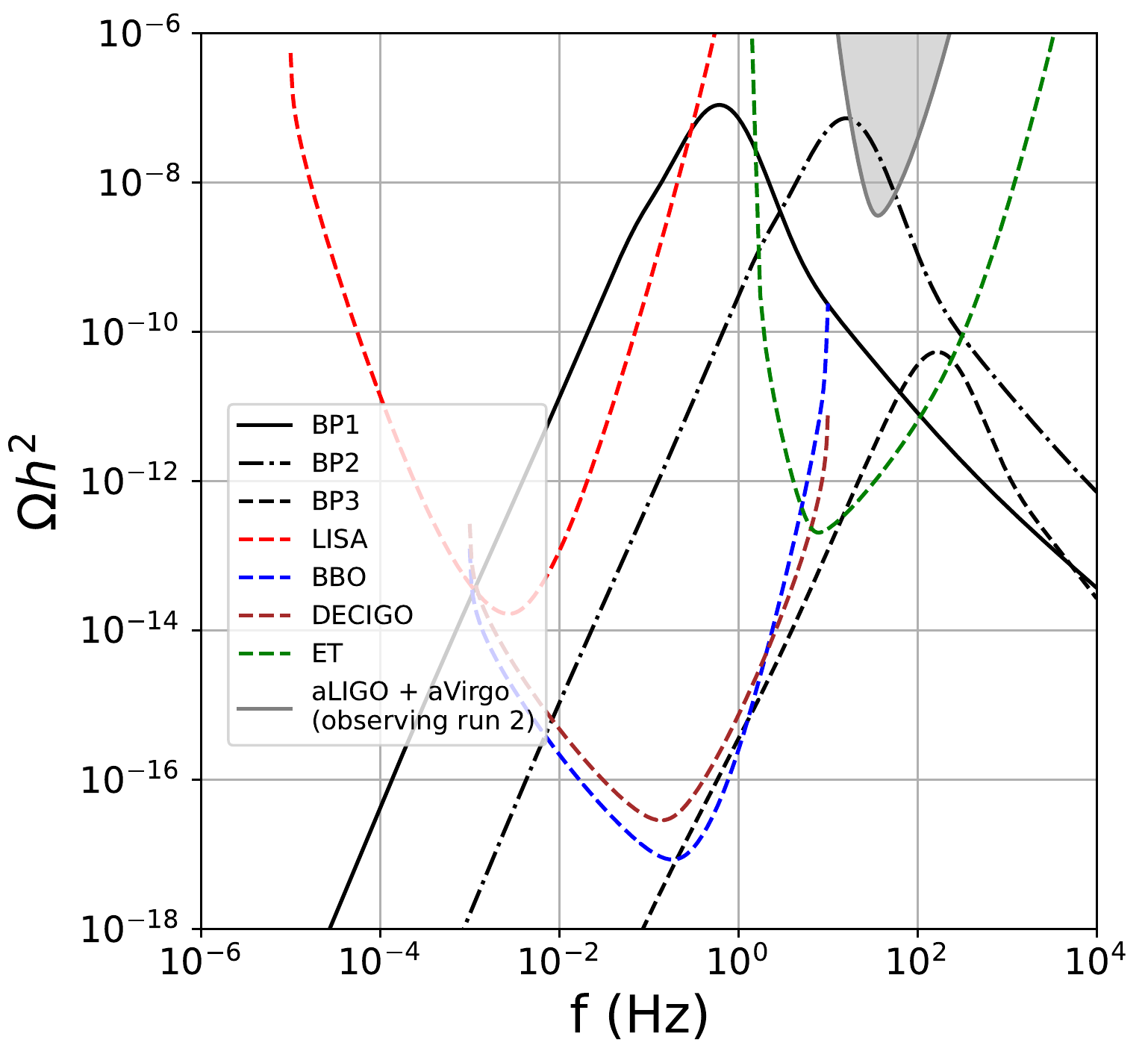} \caption{GW amplitudes for the  three benchmark points shown in Table~\ref{tab:BP} (BP1-BP3). Also shown are the current and future sensitivities of GW signal strengths.
    }
    \label{fig:BP}
\end{figure}

In Table~\ref{tab:BP}, we give three benchmark points for the $B-L$ gauge coupling strength, $B-L$ breaking scale, RHN mass scale, and the CP-phase $\phi$, which lead to successful leptogenesis. Here we have used the resonant leptogenesis mechanism~\cite{Pilaftsis:2003gt, Dev:2017wwc} so that $M_N$ in the multi-TeV range is possible, and we can get the desired CP-asymmetry $\epsilon_N$ by adjusting the CP-phase $\phi$. 
\begin{table*}[!t]
    \centering
    \begin{tabular}{|c|c|c|c|c|c|c|c|c|c|c|}\hline 
    & $\alpha_{B-L}$  & $y$ & $M$ (TeV) & $M_N$ (TeV) & $\alpha$ & $\beta/H_*$ & $v_w$ & $T_{\rm RH}$ (TeV)& $T_n$ (TeV) & $\phi$ \\ \hline 
    BP1     & 0.007 & $2.4\times 10^{-6}$ & $100$ & 7.3 & $3\times 10^6$ & 24.3 & 1 & 4.7 & 0.13 & 0.07\\
    BP2     & 0.012 & $2.9\times 10^{-6}$ & $100$ & 10 & 69 & 33.7 & 1 & 6.5 & 2.5 & $2.8\times 10^{-5}$\\
    BP3     & 0.019 & $4.4\times 10^{-6}$ & $100$ & 23 & 0.13 & 57.4 & 1 & 15 & 15 & $1.6\times 10^{-6}$\\\hline 
    \end{tabular}
    \caption{Values of the relevant phenomenological parameters giving the observed baryon asymmetry and observable gravitational wave signals for our three benchmark points.}
    \label{tab:BP}
\end{table*}
The corresponding parameters relevant for the computation of the GW amplitude, namely, $\alpha$, $\beta/H_*$, $v_w$, $T_{\rm RH}$, and $T_n$ are also given in Table~\ref{tab:BP}.  Fig.~\ref{fig:BP} shows the corresponding GW amplitudes for the three benchmark points (black curves) as a function of the GW frequency. 
Also shown in Fig.~\ref{fig:BP} are the experimental sensitivities of the current and future GW experiments, such as aLIGO~\cite{LIGOScientific:2016wof}, LISA~\cite{LISA:2017pwj}, BBO~\cite{Corbin:2005ny}, DECIGO~\cite{Musha:2017usi}, and ET~\cite{Punturo:2010zz}.  The shaded region shows the current LIGO-VIRGO exclusion~\cite{Renzini:2019vmt, KAGRA:2021kbb}, and therefore, LIGO-VIRGO is already probing part of the leptogenesis parameter space, as we will show below. 

\begin{figure*}[!ht]
    \centering
    \begin{tabular}{lcr}
    \includegraphics[width=0.33\textwidth]{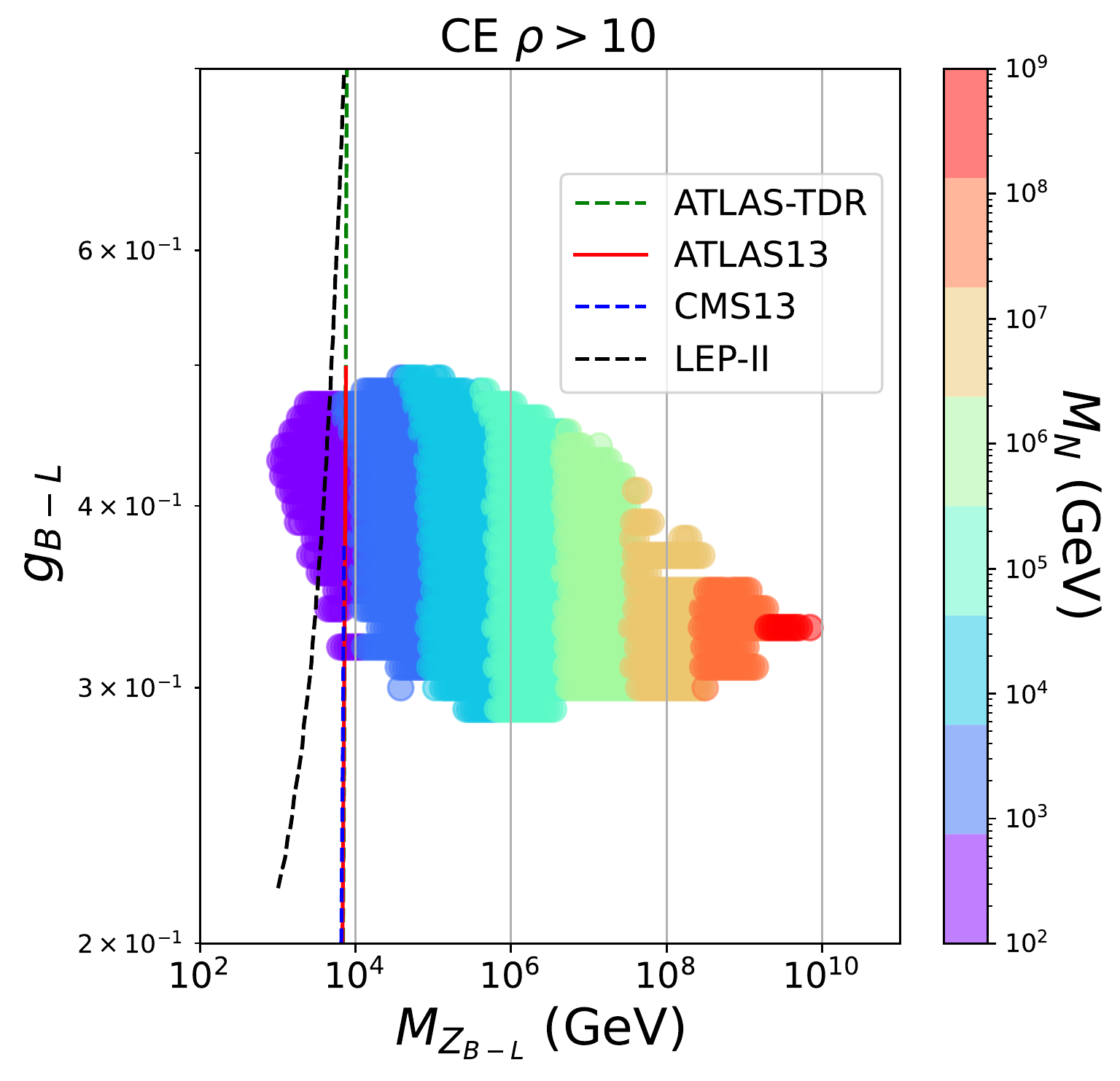} &
    \includegraphics[width=0.33\textwidth]{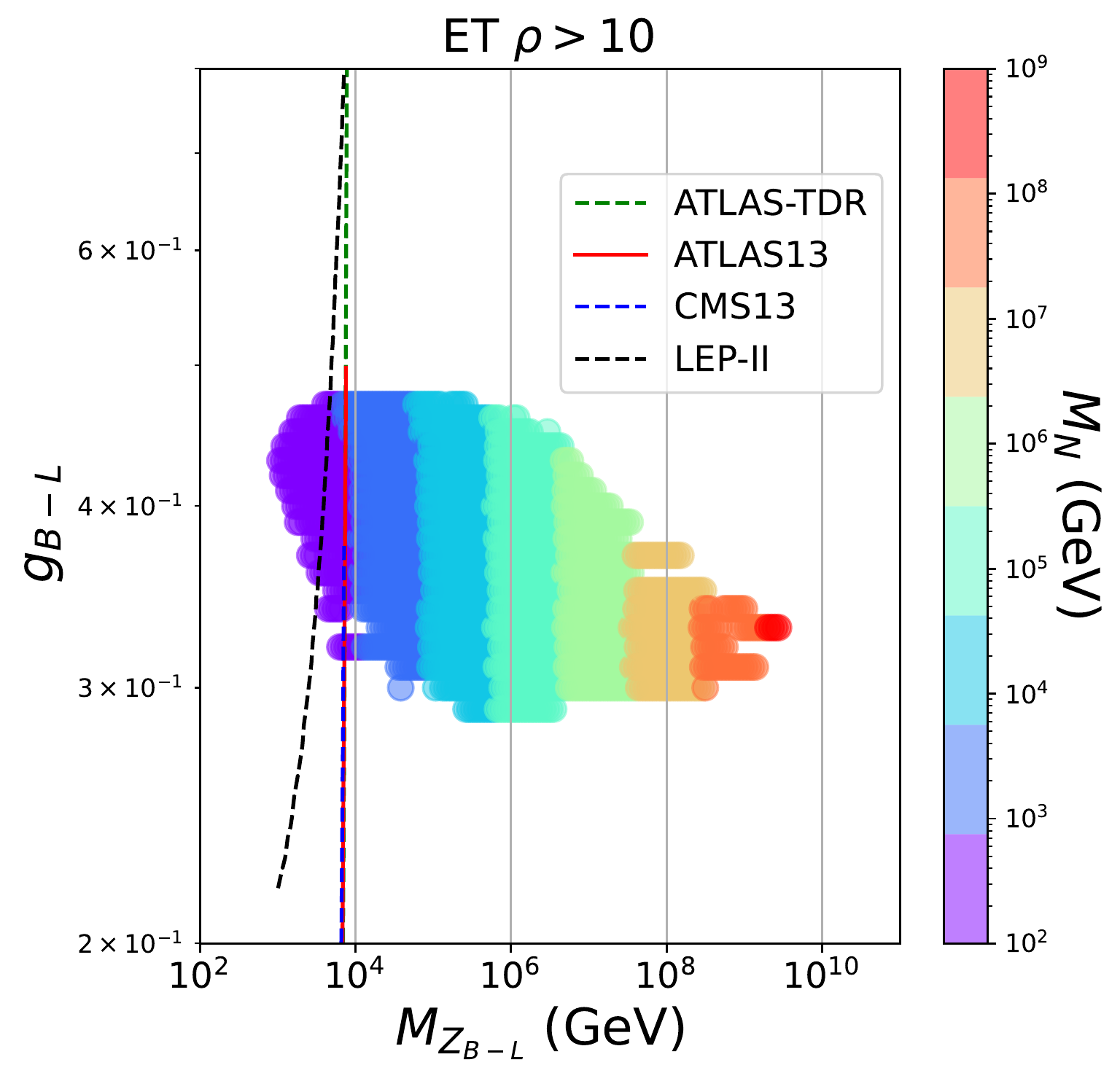} &
    \includegraphics[width=0.33\textwidth]{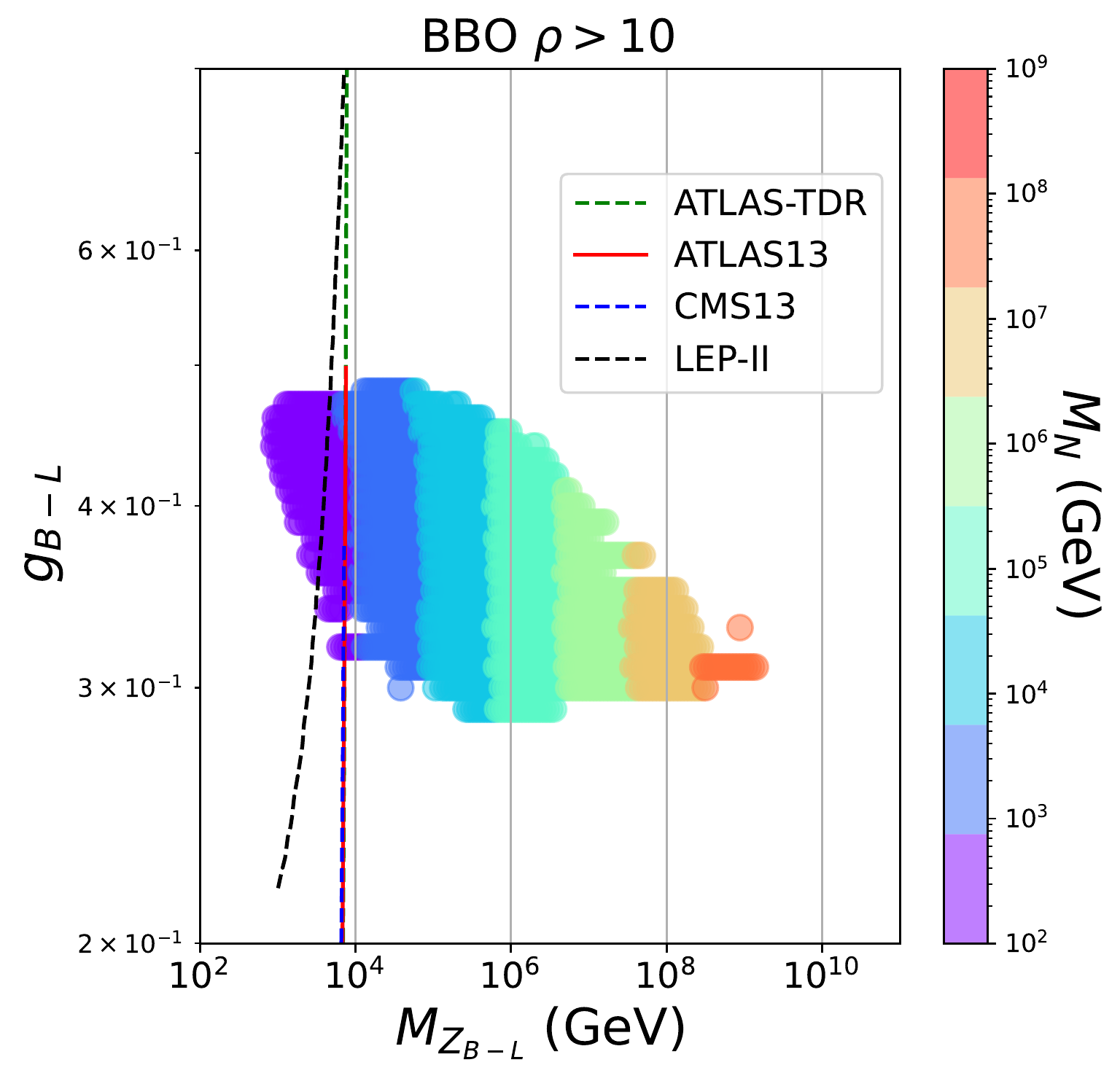} \\
    \includegraphics[width=0.33\textwidth]{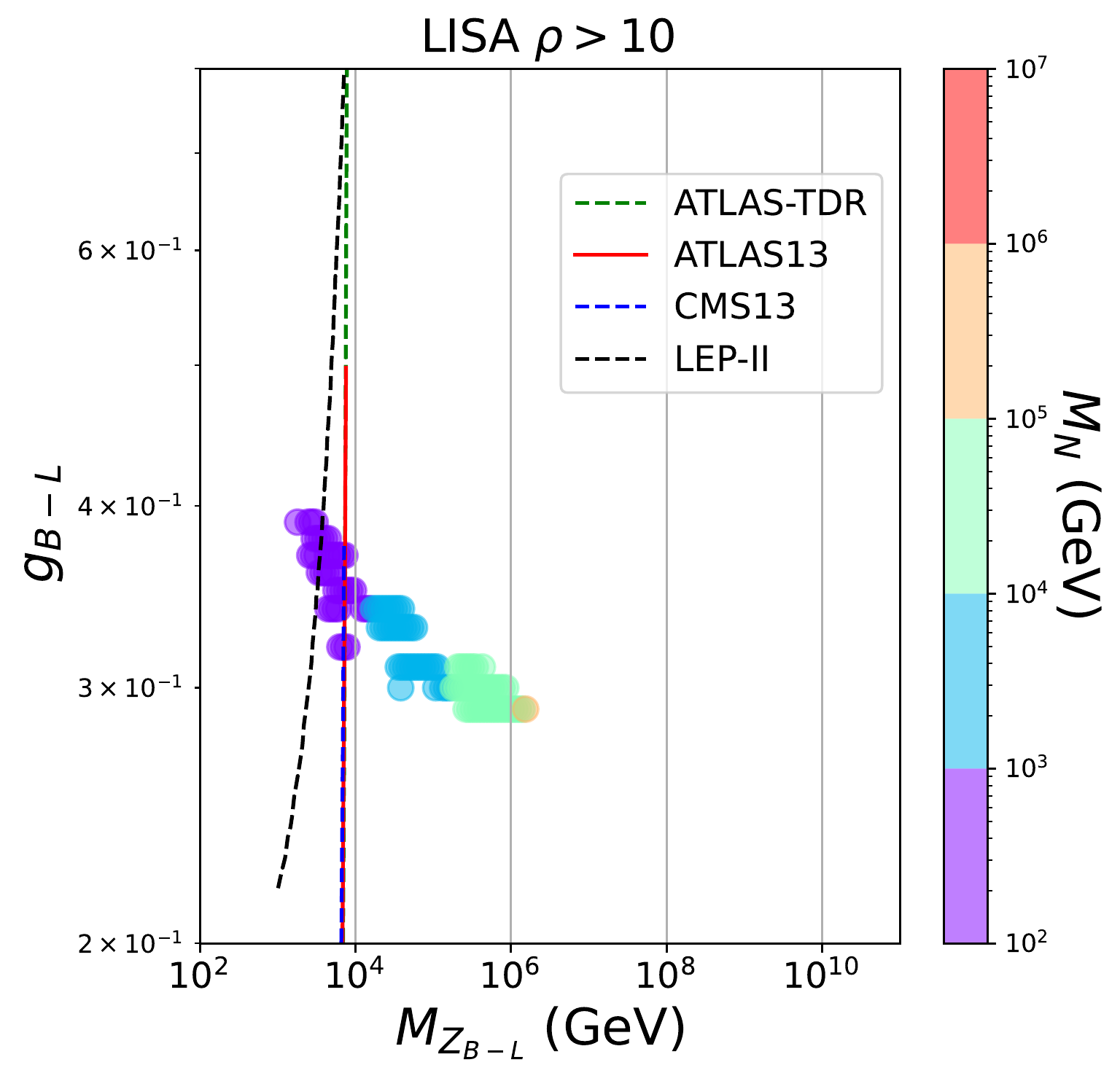} &
    \includegraphics[width=0.33\textwidth]{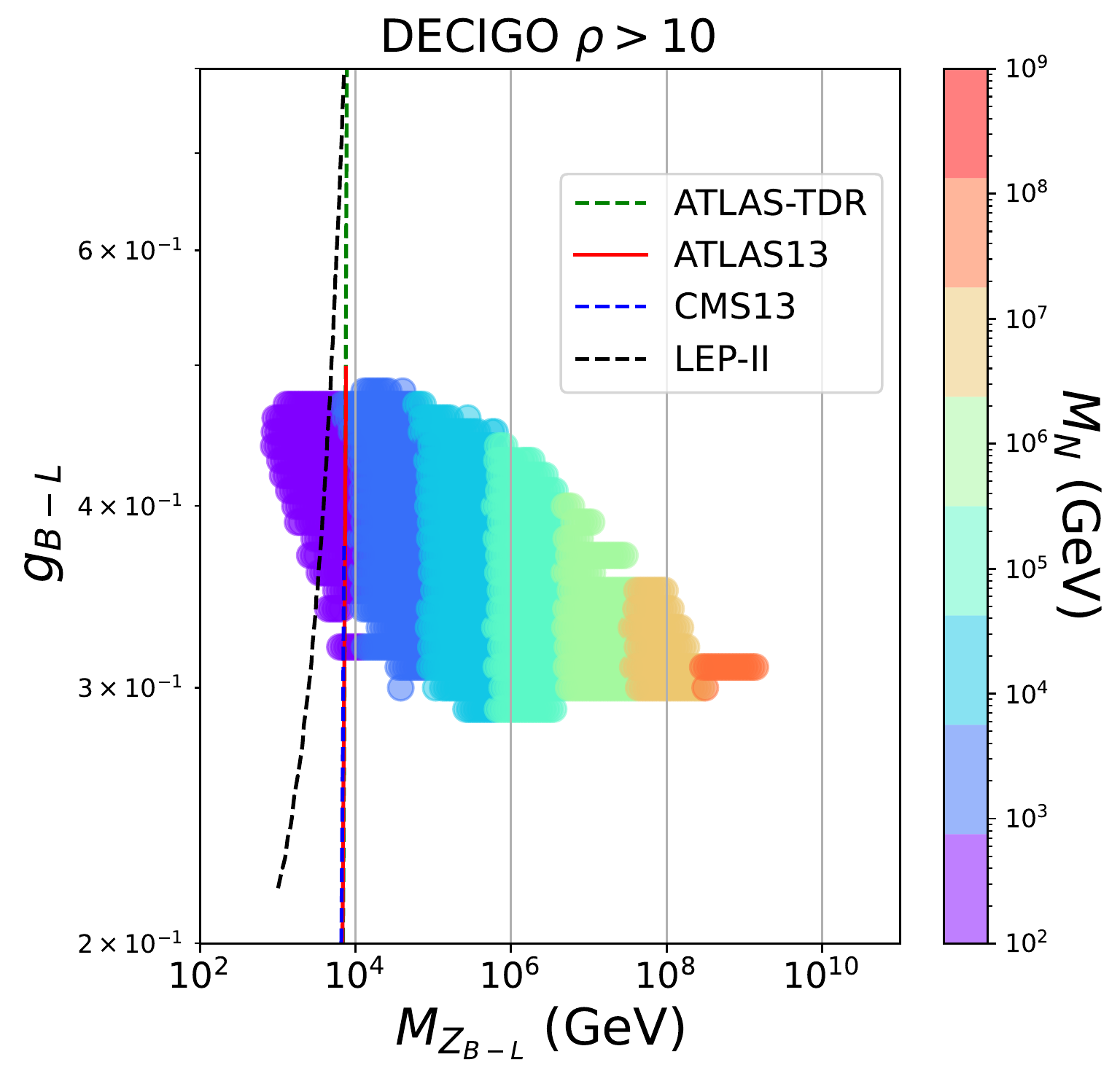} &
    \includegraphics[width=0.33\textwidth]{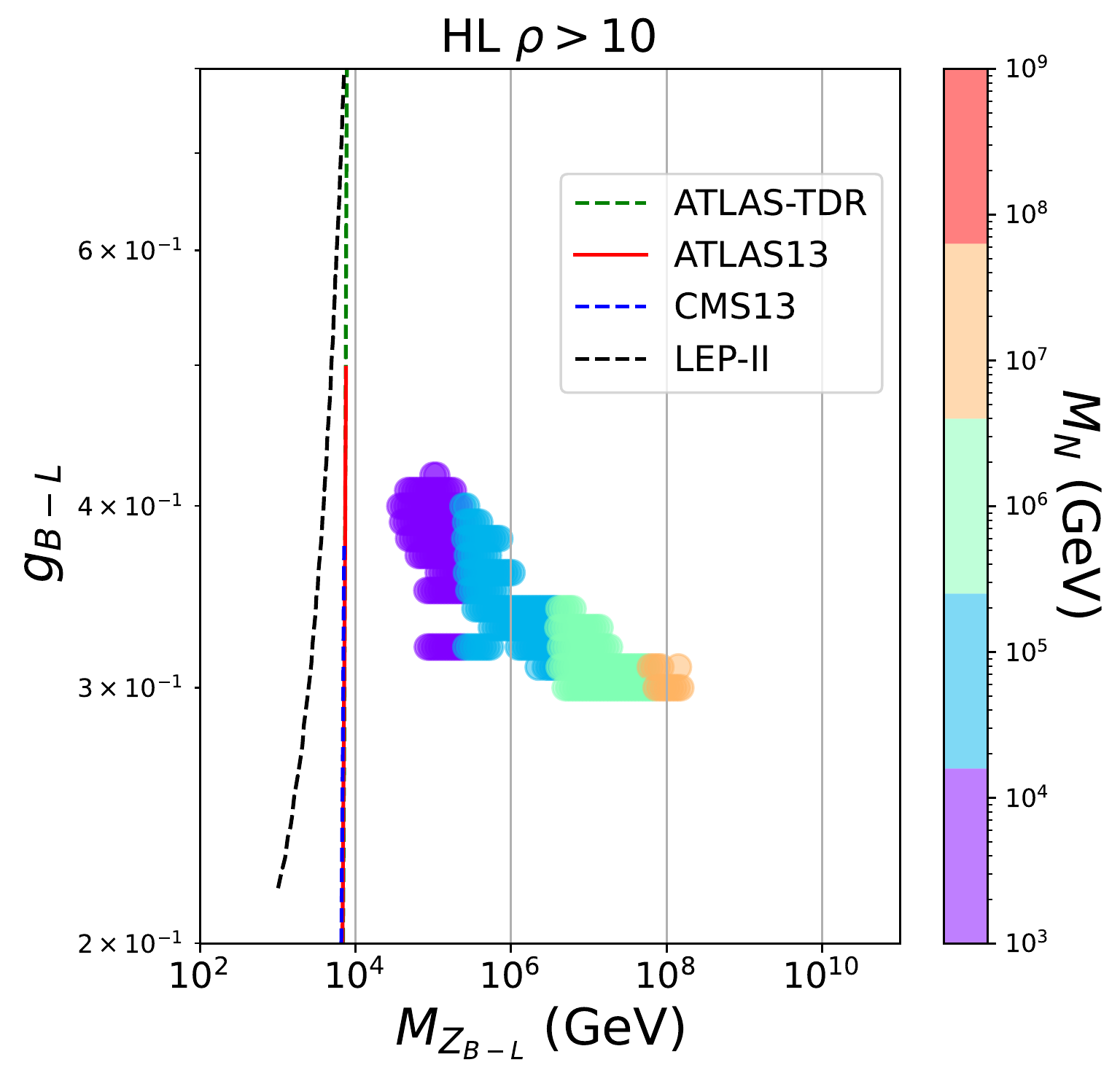} \\
    \includegraphics[width=0.33\textwidth]{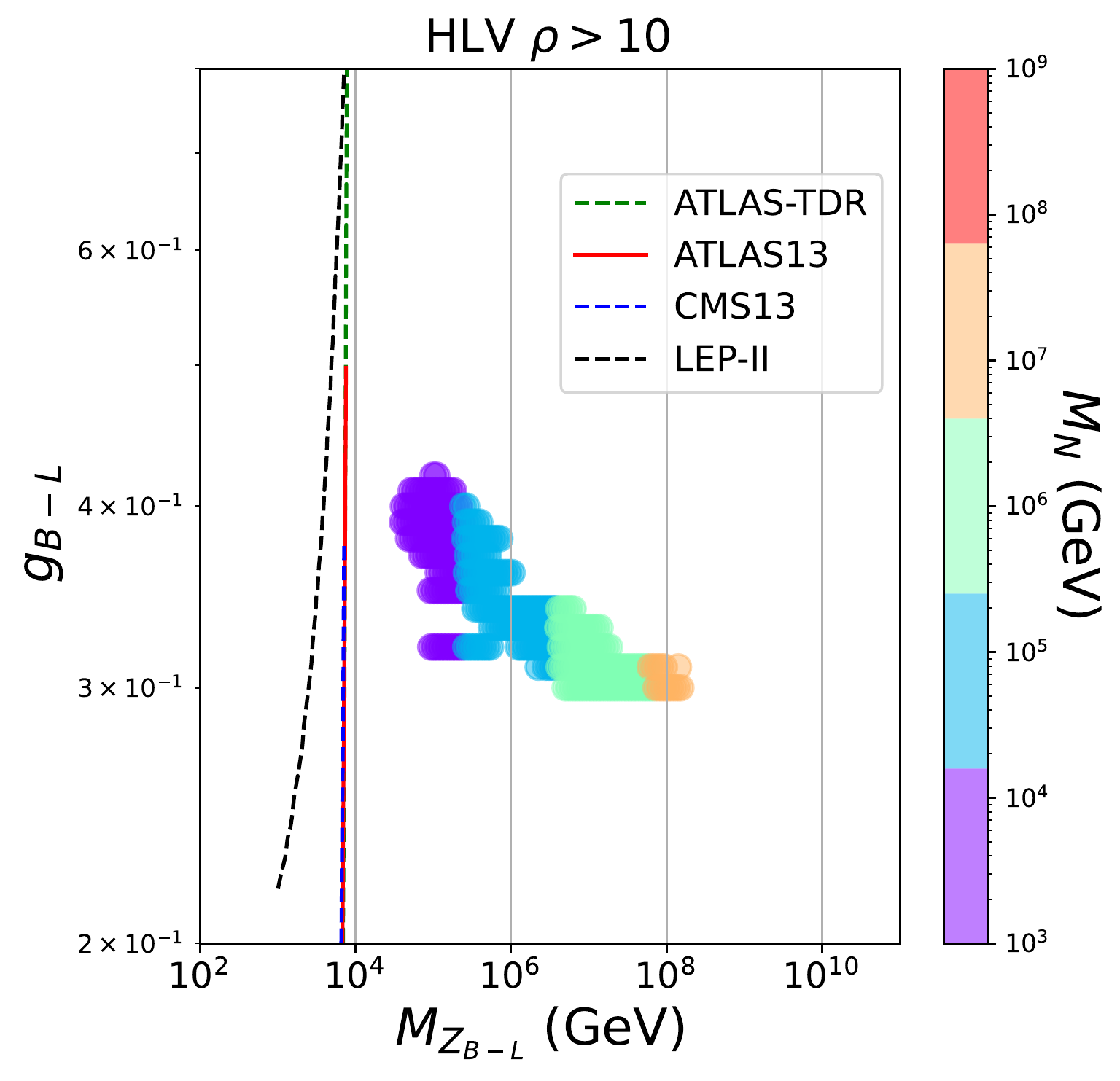} &
    \includegraphics[width=0.33\textwidth]{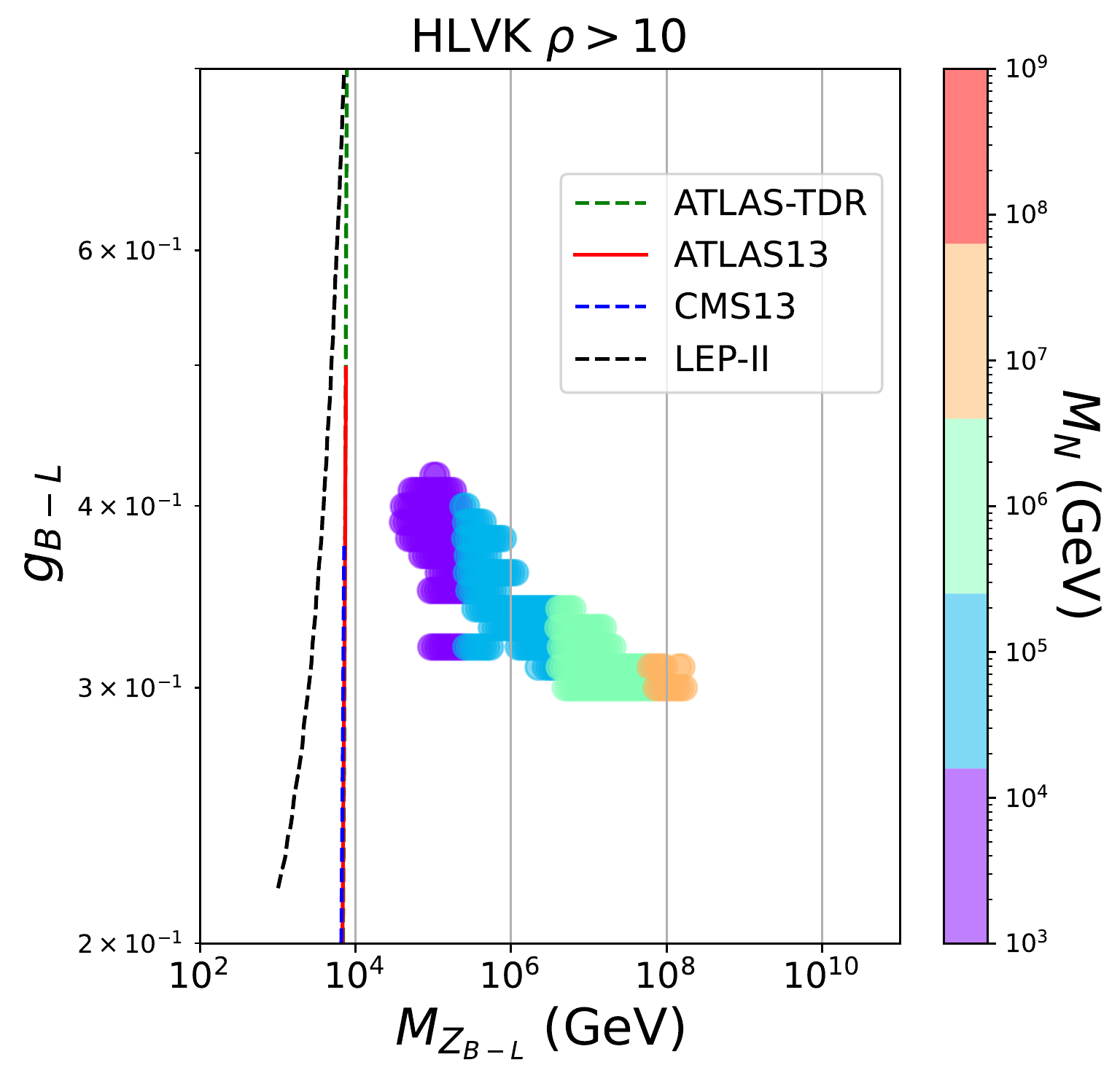} &
    \includegraphics[width=0.33\textwidth]{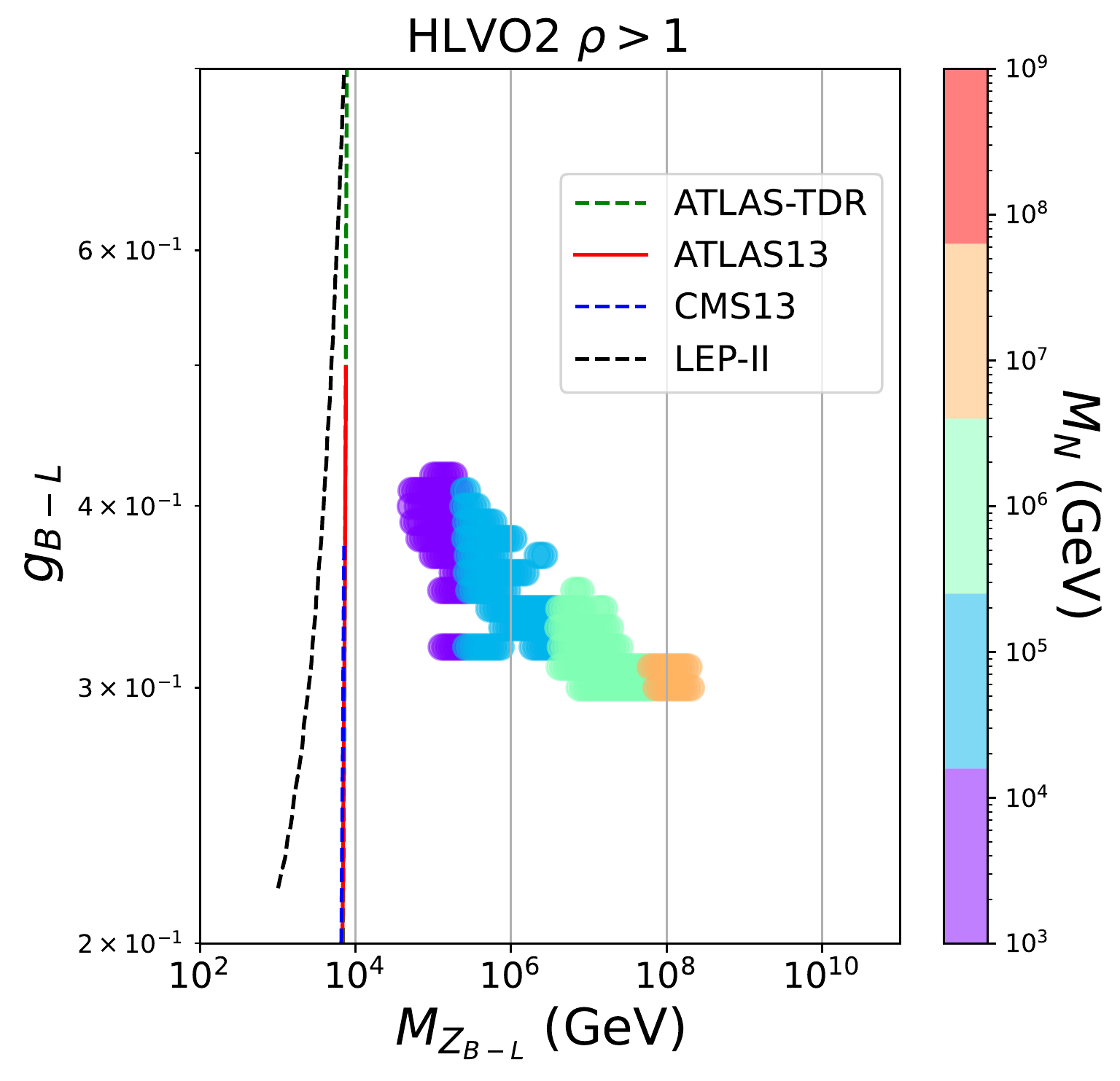}
    \end{tabular}
    \caption{The allowed parameter space for successful leptogenesis and observable GW in different GW detectors (CE, ET, BBO, LISA, DECIGO, and LIGO-VIRGO with different runs (HL-HLV02)). The dashed and solid lines show the existing (and future) collider constraints. }
    \label{fig:scan}
\end{figure*}

In order to estimate the signal strength with the ongoing GW experiments and 
also to obtain predictions for the future ones we have calculated the associated 
{\it signal-to-noise ratio} (SNR) $\rho$ by integrating over the experiment's total observing 
time $t_{\rm obs}$ and accessible frequency range $[f_{\rm min},f_{\rm max}]$~\cite{Maggiore:1999vm,Allen:1996vm,Allen:1997ad}:
\begin{equation}
\rho = \sqrt{t_{\rm obs}\,\int_{f_\text{min}}^{f_\text{max}}\,df\,\left[\frac{\Omega_\text{GW}(f)\,h^2}{\Omega_\text{expt}(f)\,h^2}\right]^2} \, ,
\end{equation}
where the running time for each experiment has been taken to be 1 year. The parameter region for $g_{B-L}$ for a given symmetry-breaking scale $M$ has four natural constraints with respect to each GW experiment, as shown by the scatter plots in Fig.~\ref{fig:scan}, assuming an SNR $\rho>10$, except for the last panel where we show $\rho>1$ for the current observing run of LIGO-VIRGO.  The upper bound on $g_{B-L}$ comes from the fact that when the coupling becomes stronger, the $\beta/H_*$ starts increasing, thus leading to the decrease of GW amplitude. It also shifts the peak frequency towards the higher values pushing the signal out-of-reach of a particular experiment. The value of $\alpha$ also decreases when the coupling is stronger. On the other hand, if one decreases the gauge coupling the ratio $T_n/T_{\rm RH} \ll 1$, resulting in the requirement of higher $\epsilon$ to satisfy Eq.~\eqref{eq:eps}. Since $\epsilon < 1$, it always forces a lower bound on $g_{B-L}$. The lower bound on the mass scale $M$ comes from the requirement of percolation temperature $T_p$ to be above the electroweak phase transition scale, i.e $T_p > v$. And finally, the higher mass scale is bounded by the sensitivity reach of each experiment as the higher mass scale corresponds to higher peak frequency. 

Also shown in Fig.~\ref{fig:scan} are the collider constraints from LEP and LHC~\cite{Das:2021esm}. It is clear that the GW sensitivities extend to higher $B-L$ breaking scale and are therefore complementary to the collider constraints. Moreover, as shown in the bottom right plot of Fig.~\ref{fig:scan}, the current LIGO-VIRGO constraint on stochastic GW~\cite{Renzini:2019vmt, KAGRA:2021kbb} has already ruled out a new portion of the $B-L$ parameter space giving rise to successful leptogenesis.   

\section{Conclusion}
\label{sec:conclusion}
Due to the availability of new tools to detect GW in the near future, probing the scale of new physics via GW has been a topic of great interest with several proposals ranging from GW sourced by cosmic strings~\cite{Dror:2019syi} and by domain walls~\cite{Barman:2022yos} to inflationary GW~\cite{Bhaumik:2022pil} being considered recently. In this paper
we proposed GW production from strong first order phase transitions in a classically scale-invariant minimal $B-L$ model as a pathway to testable resonant leptogenesis which occurs via the {\it mass-gain} mechanism~\cite{Baldes:2021vyz}. Moreover unlike the above-mentioned GW signatures, in our prescription, one is able to complement the GW signals with $Z^{\prime}$ searches in colliders (see Fig.~\ref{fig:scan}). The scale of leptogenesis that is  probed via this mechanism and its correlation with the GW signals are also different from the other scenarios as the GW spectra from various sources are different and distinguishable from each other (see Ref.~\cite{Mazumdar:2018dfl} for a review).  An interesting observation of our analysis is that the minimal requirement for the RHN Yukawa coupling $y_f$ always satisfies the washout condition at reheating temperature $T_{\rm RH}$. Furthermore, the current  LIGO-VIRGO run-3 data has already ruled out some of the parameter space for leptogenesis as shown in Fig.~\ref{fig:scan} (bottom right panel). Most interestingly, the allowed parameter regions are bounded from all sides: the lower bounds on coupling $g_{B-L}$ and $M_{Z_{B-L}}$ are coming from the requirement of $\epsilon_N < 1$ and $T_p > 100$ GeV, and the upper bounds are coming from the experimental sensitivities. In the near future the proposed GW experiments, like Einstein Telescope and  Cosmic Explorer, will have the capacity to constrain even more parameter space for thermal leptogenesis, complementary to the collider bounds. Still higher scale leptogenesis can in principle be probed by going beyond the traditional interferometer-based GW detectors to other GW detectors operating at higher frequencies; however such detectors are yet to reach suitable sensitivities although several recent proposals have been made in this direction~\cite{Aggarwal:2020olq,Berlin:2021txa, Herman:2022fau}. Finally we envisage our prescription to have discerning effects on the GW observations from electroweak phase transitions (typically in LISA) if the leptogenesis occurs very close to the EW scale~\cite{Pilaftsis:2005rv} but this study is beyond the scope of the current paper and will be taken up in future.


\acknowledgments
The work of B.D. is supported in part by the US Department of Energy under Grant No.~DE-SC0017987. A.M.’s research is funded by the Netherlands Organisation for Science and Research (NWO) grant number 680-91-119.

\section*{Note Added} During the final stages of the writing of this manuscript, we noticed Ref.~\cite{Huang:2022vkf} which uses the same mechanism, but focuses on the high-scale leptogenesis. 

\bibliographystyle{apsrev4-1}

\bibliography{ref}
\end{document}